# Nanoscale $Sr_2IrO_4$ Freestanding Thin-Films for Flexible Electronics

Sujan Shrestha[†], Matthew Coile[†], Menglin Zhu[‡], Maryam Souri[†], Jiwoong Kim[†,§], Rina Pandey[†], Joseph Brill[†], Jinwoo Hwang[‡], Jong-Woo Kim[#], and Ambrose Seo[*,†]

[†]*Department of Physics and Astronomy, University of Kentucky, Lexington, KY 40506, USA*
[‡]*Department of Materials Science and Engineering, The Ohio State University, Columbus, Ohio 43210, USA*
[§]*Department of Physics, Pusan National University, Busan 46241, South Korea*
[#]*Advanced Photon Source, Argonne National Laboratory, Argonne, Illinois 60439, USA*

ABSTRACT: We report the structural and optical properties of nanoscale $Sr_2IrO_4$ freestanding thin-films fabricated using a water-soluble $Sr_3Al_2O_6$ layer. The coherent lattice structure, phonon modes, two-magnon Raman scattering, and optical absorption spectra of the $Sr_2IrO_4$ nanomembrane are analogous to those of the layered iridate epitaxial thin-films and single crystals. Remarkably, the formation of 3-unit-cell-thick $SrIrO_3$ and interfacial composite layers alleviates antiphase boundaries at the $Sr_2IrO_4$/$Sr_3Al_2O_6$ interface, resulting in structurally-robust nanomembranes. Our experimental results show that this freestanding thin-film approach of layered oxides can provide techniques for tuning or realizing unprecedented states beyond conventional thin-film methods, suggesting a pathway in achieving flexible layered-oxide electronics.

Recently, freestanding metal-oxide thin-films have attracted attention due to their tunability and potential integration with flexible electronic devices. Constraints imposed by the substrate, such as epitaxial strain and interfacial dislocation, do not play a role in the stand-alone thin-films. Therefore, beyond thermodynamic equilibrium, the mechanical modification of freestanding transition-metal oxide (TMO) thin-films can provide an unprecedented state because their properties are due to strong interactions between charge, spin, and orbital degrees of freedom.[1,2] Hence, freestanding TMO thin-films are an ideal system for manipulating and realizing novel devices such as flexible electronics,[1,3-5] ferroelectric memory devices,[6,7] and flexible energy conversion devices[8] by integrating them with semiconductor devices or flexible platforms.

Maintaining the high crystalline quality of freestanding TMO thin-film is essential not only for their functional properties but also for device applications. Recently, Lu *et al.* were able to synthesize high-quality freestanding $SrTiO_3$ and $La_{0.7}Sr_{0.3}MnO_3$ thin-films using a water-soluble $Sr_3Al_2O_6$ buffer layer.[9] Note that the perovskites (i.e., $AMO_3$, where $A$ is an alkaline earth or rare earth element and $M$ is a transition-metal element) and $Sr_3Al_2O_6$ have great structural compatibility. However, it is unknown whether the same $Sr_3Al_2O_6$ buffer layer will also work for synthesizing freestanding *layered* oxide (e.g., $K_2NiF_4$-type structure) thin-films. While the in-plane square lattice of layered oxides can match well with that of $Sr_3Al_2O_6$, the large difference in the $c$-axis lattice constant can result in the nucleation of antiphase boundaries[10] near the step-terrace edges of the substrate. This difference can severely deteriorate the structural integrity of the freestanding layered oxide thin-films and interfere with its transport to an intended device.

In this letter, we report that 24-nm-thick freestanding thin-films of epitaxial $Sr_2IrO_4$ can be synthesized using pulsed laser deposition and water-leaching process. $Sr_2IrO_4$ is a relativistic Mott insulator with the exotic $J_{eff}$= 1/2 pseudo-spin due to the confluence of strong spin-orbit interaction and electron correlation.[11,12] Due to its structural and magnetic similarities with high-$T_c$ cuprates, this compound has attracted much attention recently.[13-16] We separated a freestanding $Sr_2IrO_4$ thin-film from its substrate by dissolving a water-soluble $Sr_3Al_2O_6$ buffer layer, following the description of Ref. [8]. We confirmed the high-crystallinity and structural integrity of our freestanding $Sr_2IrO_4$ thin-films, whose quality is comparable to that of epitaxial thin-films and single crystals, using various characterization methods. Our results extend the scope of freestanding thin-films to the fundamental research and device application of layered oxides such as high-$T_c$ cuprate superconductors.

We grew epitaxial heterostructures of $Sr_2IrO_4/Sr_3Al_2O_6$ on $SrTiO_3$ (001) substrates using pulsed laser deposition (PLD). Atomically flat surfaces of $SrTiO_3$ (001) substrates were prepared by deionized-water leaching and thermal annealing, as described in Ref. [17]. We deposited $Sr_3Al_2O_6$ layer first and then $Sr_2IrO_4$ epitaxial thin-film with $K_2NiF_4$ type crystal. To stabilize the structure of a $Sr_2IrO_4$ thin-film and facilitate its transport process, we attached a polyethylene terephthalate (PET) film on the sample surface. Following the process described in Ref. [9], we exfoliated freestanding $Sr_2IrO_4$ thin-film by dissolving the $Sr_3Al_2O_6$ buffer layer in deionized water. Figure 2(a) shows a photograph of a freestanding $Sr_2IrO_4$ nanomembrane with the entire area of 5×5 $mm^2$. After this process, we could transfer freestanding thin-films into various substrates such as silicon wafer, glass, etc., as shown in Fig. S1.

Scanning transmission electron microscopy (STEM) shows that there is a 3-unit-cell-thick $SrIrO_3$ interfacial layer between $Sr_3Al_2O_6$ and $Sr_2IrO_4$ layers. Figure 1(a) shows a cross-sectional *Z*-contrast STEM image of an as-grown $Sr_2IrO_4/Sr_3Al_2O_6/SrTiO_3$ sample. High-resolution data of the $Sr_2IrO_4/Sr_3Al_2O_6$ interface (Fig. 1(b)) reveal that there are three unit-cell of $SrIrO_3$ and an interfacial composite layer between $Sr_2IrO_4$ and $Sr_3Al_2O_6$ while a sharp interface exists between $Sr_3Al_2O_6$ and $SrTiO_3$ (Fig. 1(c)). Similar interfacial perovskite layers have been observed in various heterointerfaces of complex oxide thin-films synthesized by molecular beam epitaxy (MBE) and PLD.[18-21] Note that the interfacial composite layer, which is formed presumably due to the intermixing of $SrIrO_3$ and $Sr_3Al_2O_6$ layers, is only about two-monolayer-thick. Such an atomic-scale interfacial layer has also been observed in the heterointerfaces of $KTaO_3/GdScO_3$ (Ref. 22) and $(La_{0.3}Sr_{0.7})(Al_{0.65}Ta_{0.35})O_3/SrTiO_3$ (Ref. 23). These interfacial perovskite and composite layers seem necessary for the thermodynamical stability in many complex oxide heterointerfaces. It is noteworthy that the interfacial composite and 3-unit-cell-thick $SrIrO_3$ layers prevent the formation of any antiphase boundaries in the $Sr_2IrO_4$ layer, resulting in the high-crystallinity of freestanding $Sr_2IrO_4$ membranes.

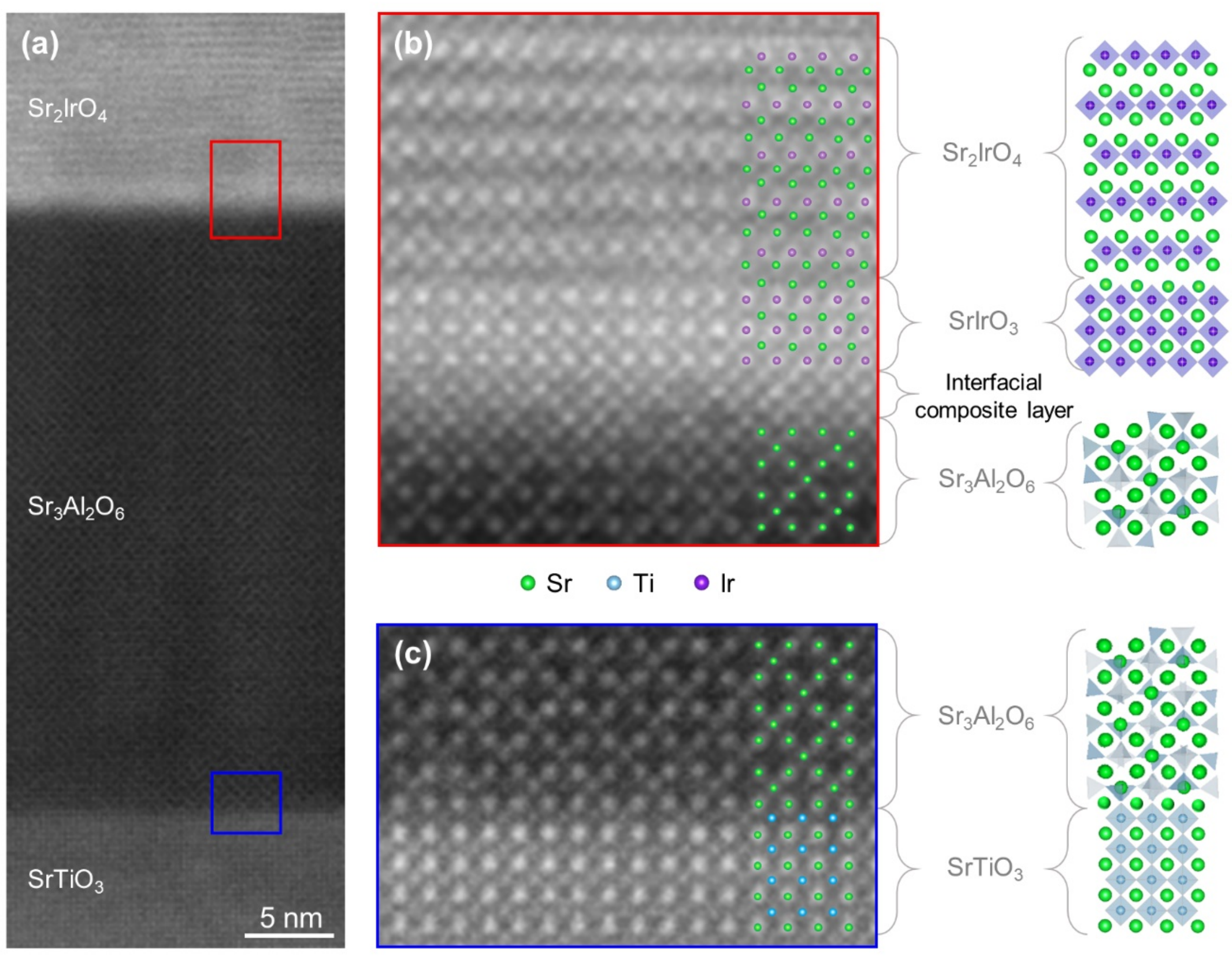

**Fig. 1. (a)** A cross-sectional *Z*-contrast STEM image of an as-grown $Sr_2IrO_4/Sr_3Al_2O_6$ bilayer on a $SrTiO_3$ (001) substrate. High-resolution STEM images of **(b)** $Sr_2IrO_4/Sr_3Al_2O_6$ (red box) and **(c)** $Sr_3Al_2O_6/SrTiO_3$ (blue box) interfaces, respectively. The schematic diagram shows microscopic structures obtained from the intensity profiles proportional to the atomic number (*Z*) of each ion. Note that there are interfacial composite layer and three layers of $SrIrO_3$ at the $Sr_2IrO_4/Sr_3Al_2O_6$ interface. Nevertheless, PLD growth conditions can stabilize epitaxial $Sr_2IrO_4$ layers on top of the interface.

X-ray diffraction confirms that the structure of the freestanding $Sr_2IrO_4$ thin-film is coherent. Figure 2(b) shows the $2\theta$-$\theta$ scans of a 24-nm-thick freestanding $Sr_2IrO_4$ membrane placed on a silicon wafer and a $Sr_2IrO_4/Sr_3Al_2O_6/SrTiO_3$ sample. Well-defined (*00l*) peaks from $Sr_2IrO_4$ are presented in both the samples, while the freestanding $Sr_2IrO_4$ thin-film does not show any peaks from $Sr_3Al_2O_6$ or $SrTiO_3$. However, the broad full-width half-maximum of the rocking curve (Fig. 2(c)) indicates that the freestanding $Sr_2IrO_4$ thin-film has more mosaic spread than the as-grown $Sr_2IrO_4$ thin-film on $Sr_3Al_2O_6/SrTiO_3$, presumably due to structural degradation during the exfoliation and transfer processes.[24] Nevertheless, the single peak rocking curve indicates that multiple domains are not present. We obtained the in-plane and out-of-plane lattice parameters from X-ray reciprocal space mapping (RSM), as shown in Fig. 2(d), near the 103-reflection of $SrTiO_3$ substrates. Clear $11\underline{18}$- and $40\underline{12}$-reflections are visible from $Sr_2IrO_4$ and $Sr_3Al_2O_6$, respectively. From these RSM data, we can see that the as-grown $Sr_2IrO_4$ thin-film is under tensile strain although the $Sr_3Al_2O_6$ buffer layer is partially relaxed. The freestanding $Sr_2IrO_4$ thin-film has longer *c*-axis lattice constant than the as-grown $Sr_2IrO_4$ thin-film while their in-plane lattice constants are similar to each other. Table 1 shows the in-plane and out-of-plane lattice constants and strain values $\varepsilon_{xx}$ and $\varepsilon_{zz}$. Note that the freestanding $Sr_2IrO_4$ thin-film is under tensile strain, i.e., larger in-plane lattice and smaller out-of-plane lattice than $Sr_2IrO_4$ single crystals, presumably due to a small number of defects such as oxygen vacancies created during the sample synthesis.

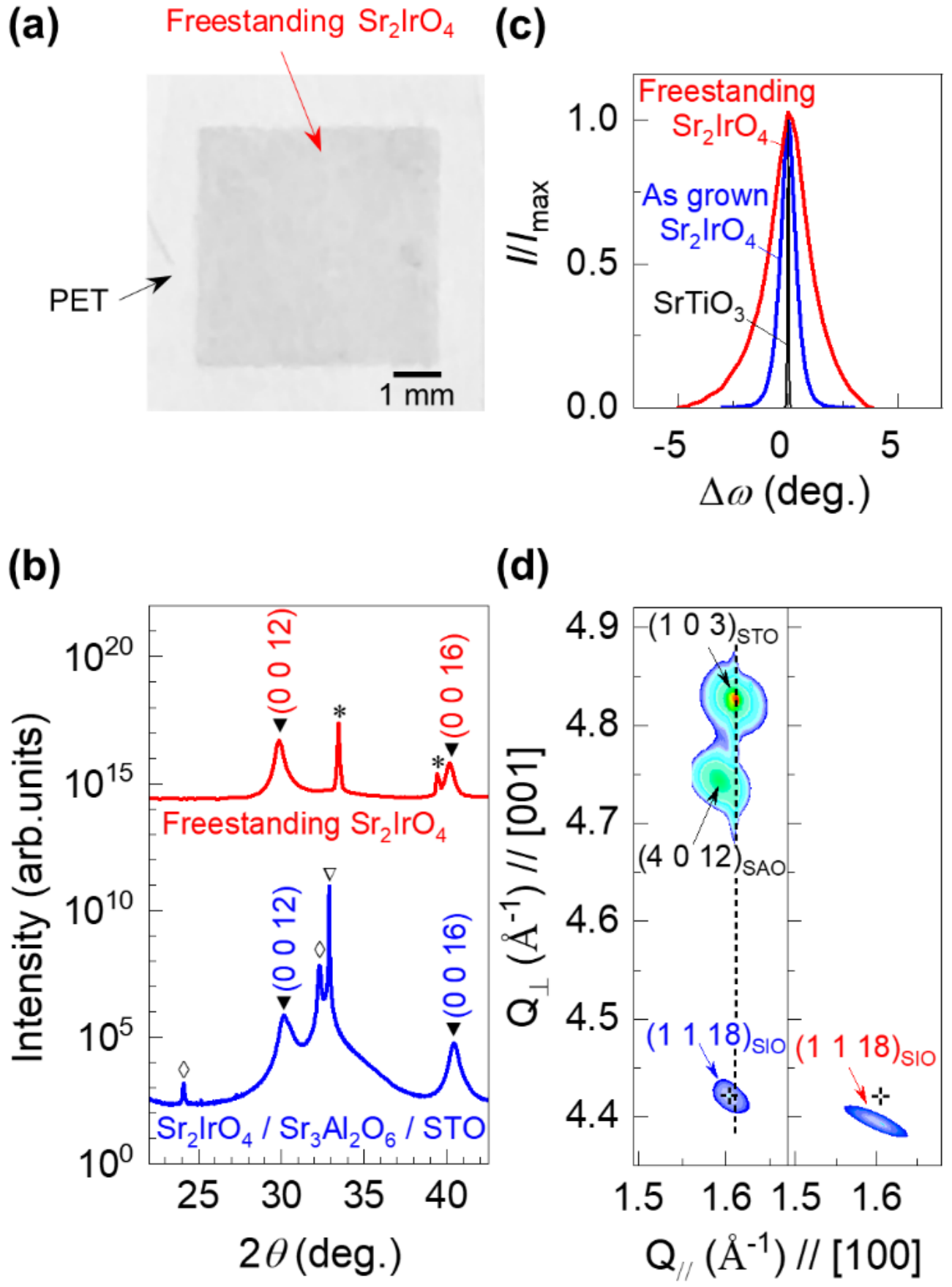

**Fig. 2. (a)** A photograph of a freestanding $Sr_2IrO_4$ nanomembrane transferred on a polyethylene terephthalate (PET) films. **(b)** X-ray diffraction *2θ-θ* scans of a $Sr_2IrO_4/Sr_3Al_2O_6/SrTiO_3$ sample and a freestanding $Sr_2IrO_4$ membrane transferred to a silicon wafer. The symbols [▼] and [◊] indicate the peaks from the $Sr_2IrO_4$ thin-film and the $Sr_3Al_2O_6$ layer, respectively. The asterisks [*] and symbol [∇] indicate peaks from the substrates, i.e., silicon and $SrTiO_3$, respectively. **(c)** X-ray rocking curves of a freestanding $Sr_2IrO_4$ thin-film on silicon and an as-grown

$Sr_2IrO_4/Sr_3Al_2O_6/SrTiO_3$. The increased FWHM of the freestanding $Sr_2IrO_4$ thin-film (2.11°) compared to the as-grown $Sr_2IrO_4$ thin-film (0.84°) implies that mosaicity increases during the transfer process. **(d)** X-ray reciprocal space mapping of as-grown $Sr_2IrO_4/Sr_3Al_2O_6/SrTiO_3$ and a freestanding $Sr_2IrO_4$ thin-film near the 103-reflection of $SrTiO_3$. Both samples clearly show the $11\underline{18}$-reflection of $Sr_2IrO_4$. The cross symbol indicates the peak position of the as-grown $Sr_2IrO_4/Sr_3Al_2O_6/SrTiO_3$ for comparison with the freestanding $Sr_2IrO_4$ thin-film.

**Table 1.** In-plane and out-of-plane lattice constants and strain values of as-grown $Sr_2IrO_4$ and freestanding $Sr_2IrO_4$ thin-films.

| | $a$ (Å) | $c$ (Å) | $\varepsilon_{xx}$ (%) | $\varepsilon_{zz}$ (%) | Poisson's ratio |
|---|---|---|---|---|---|
| As-grown $Sr_2IrO_4$ | 5.54 | 25.50 | +0.73 (±0.30) | -1.12 (±0.16) | 0.49 (±0.20) |
| Freestanding $Sr_2IrO_4$ | 5.53 | 25.74 | +0.55 (± 0.35) | -0.18 (± 0.06) | 0.26 (±0.20) |

$\varepsilon_{xx} = (a_{\text{film}} - a_{\text{bulk}}) / a_{\text{bulk}} \times 100$ (%) and $\varepsilon_{zz} = (c_{\text{film}} - c_{\text{bulk}}) / c_{\text{bulk}} \times 100$ (%)

Polarization-dependent Raman spectroscopy shows that the lattice dynamics of the freestanding $Sr_2IrO_4$ membrane is consistent with that of $Sr_2IrO_4$ epitaxial thin-films and single crystals. Figure 3(a) shows Raman spectra of a freestanding $Sr_2IrO_4$ thin-film on a PET film measured at room temperature with $z(x'y')\bar{z}$ ($B_{1g}$), $z(x'x')\bar{z}$ ($A_{1g}$+$B_{2g}$), $z(xy)\bar{z}$ ($B_{2g}$), and $z(xx)\bar{z}$ ($A_{1g}$+$B_{1g}$) scattering geometries. The axes $x'$ and $y'$ are directed along the Ir-O bond direction and $x$ and $y$ axes are rotated 45° relative to them. The observed phonon modes are consistent with those of $Sr_2IrO_4/SrTiO_3$ thin-films and $Sr_2IrO_4$ single crystals, reported in Refs. [25-27]. Figure 3(b) shows the temperature-dependent $B_{2g}$ mode spectra of a freestanding $Sr_2IrO_4$ membrane on a silicon wafer. At low temperature, a broad two-magnon scattering peak emerges and its peak energy ($\omega_{2M}$) is approximately 1330 $cm^{-1}$ at 10 K. The overall spectral shape of the two-magnon scattering is consistent with that of $Sr_2IrO_4$ epitaxial thin-films and single crystals.[25,28] The measured $\omega_{2M}$ value implies that the exchange interaction ($J$) of the $J_{eff}$ = 1/2 pseudospins in our freestanding $Sr_2IrO_4$ thin-film is approximately 60 meV.

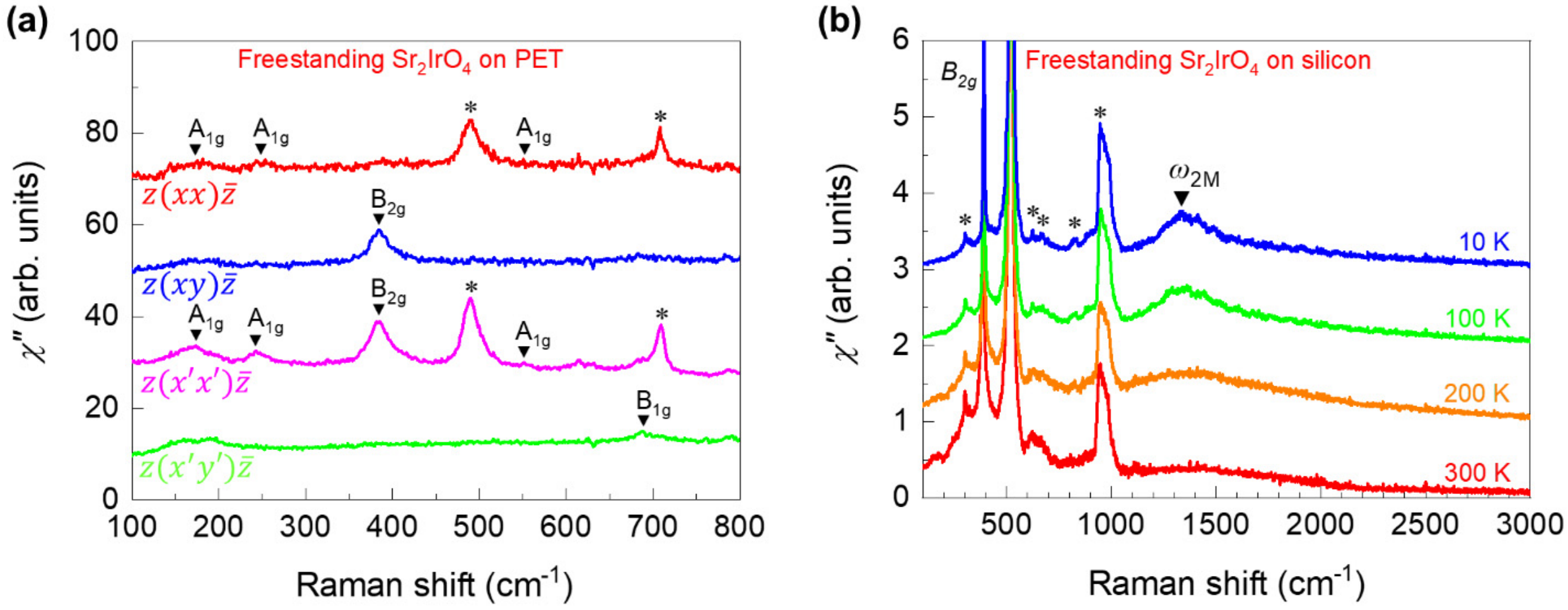

**Fig. 3. (a)** Polarization-dependent Raman spectra of a 24-nm-thick freestanding $Sr_2IrO_4$ membrane sitting on a polyethylene terephthalate (PET) membrane at room temperature. Four polarization-dependent backscattering channels $z(xx)\bar{z}$, $z(xy)\bar{z}$, $z(x'x')\bar{z}$ , and $z(x'y')\bar{z}$ probe $A_{1g}$+$B_{1g}$, $B_{2g}$, $A_{1g}$+$B_{2g}$, and $B_{1g}$ modes, respectively. The polarization-dependent phonon modes of the freestanding $Sr_2IrO_4$ thin-film is consistent with epitaxial thin-films and single crystals of $Sr_2IrO_4$, Refs. [25-27]. The asterisks [*] indicate molecular vibrations from PET. **(b)** Temperature-dependent $B_{2g}$ mode Raman spectra of a freestanding $Sr_2IrO_4$ thin-film transferred on a silicon wafer. The symbol [▼] (1335 $cm^{-1}$) indicates a two-magnon scattering peak of the $Sr_2IrO_4$

freestanding thin-film. The sharp peak around 395 $cm^{-1}$ is a $B_{2g}$ phonon of $Sr_2IrO_4$. The asterisks [*] indicate phonon modes of silicon.

From optical spectroscopic measurements (Fig. 4), we observed a blue-shifted $J_{eff}$ = 1/2 inter-band optical transition, implying that slight tensile strain exists in our freestanding $Sr_2IrO_4$ thin-film. Both as-grown and freestanding $Sr_2IrO_4$ thin-film show a well-known two-peak structure, marked as $\alpha$ and $\beta$, in their optical absorption spectra, which is consistent with the previously reported data of $Sr_2IrO_4$ thin-films and single crystals.[29-31] The overall optical absorption in the photon energies below 2 eV is due to the inter-band optical transitions from the lower Hubbard band (LHB) to the upper Hubbard band (UHB) of the $J_{eff}$ = 1/2 state.[32] Note that the optical absorption spectrum of the freestanding $Sr_2IrO_4$ thin-film is broader and shifted slightly to higher energies than the $Sr_2IrO_4/SrTiO_3$ sample reported in Ref. [29]. This observation is consistent with the X-ray diffraction data (Fig. 2) because the broad and blue-shifted optical absorption spectrum is relevant to the increased Ir-O-Ir bond angle of the freestanding $Sr_2IrO_4$ thin-films.

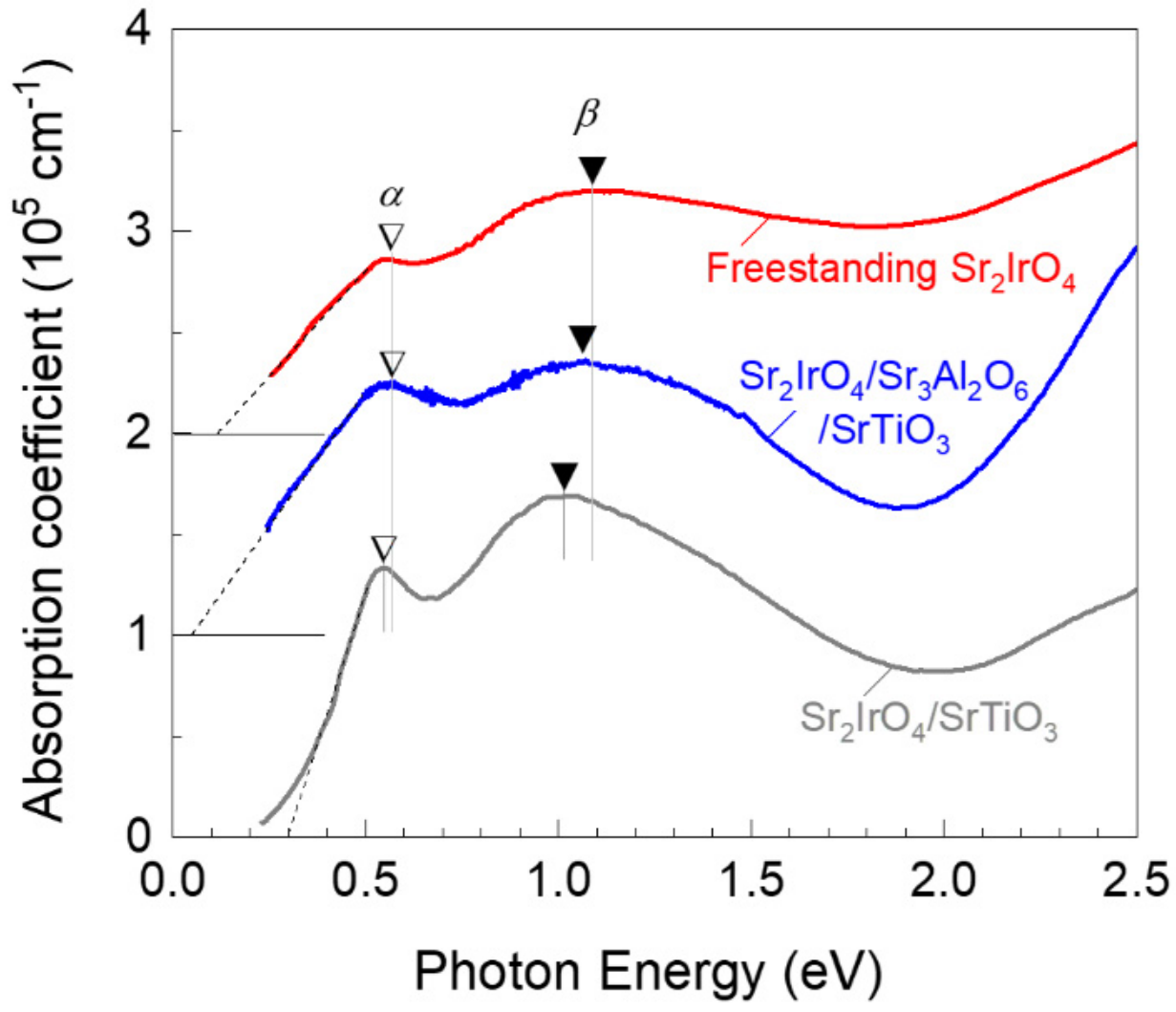

**Fig. 4.** Optical absorption coefficient spectra of freestanding $Sr_2IrO_4$ thin-films and $Sr_2IrO_4/Sr_3Al_2O_6/SrTiO_3$ samples. The spectrum of an epitaxial $Sr_2IrO_4$ thin-film grown on a $SrTiO_3$ substrate is show for comparison from Ref [29]. The spectra are vertically shifted for clarity.

In conclusion, we synthesized high-quality freestanding $Sr_2IrO_4$ nanomembranes using water-soluble $Sr_3Al_2O_6$ and conducted structural and optical characterizations of the samples. We observed coherent lattice structure, phonons modes, two-magnon Raman scattering, and optical absorption spectra of the freestanding $Sr_2IrO_4$ thin-film. While there is a slight degradation of crystallinity and generic tensile strain, the overall properties of the freestanding $Sr_2IrO_4$ thin-film are analogous to those of the layered iridate epitaxial thin-films and single crystals. Note that 3-unit-cell-thick $SrIrO_3$ and interfacial composite layers play an important role in avoiding the formation of antiphase boundaries at the heterointerface between the layered iridate and $Sr_3Al_2O_6$ and protect the structural integrity of the freestanding $Sr_2IrO_4$ thin-film. We suggest that freestanding thin-films of layered oxides with strongly correlated electrons can be a model system for tuning their physical properties beyond the thermodynamic limit of materials by combining them with piezoelectric or mechanical strain devices. Our results extend the scope of freestanding thin-films to the fundamental research beyond conventional thin-films and device application of layered oxides by integrating them with the semiconductor devices and flexible platforms.

**ASSOCIATED CONTENT**

**Supporting Information**

[Synthesis and characterization, schematic diagrams of the freestanding $Sr_2IrO_4$ thin-film, and optical transmission spectra of a freestanding $Sr_2IrO_4$ thin film on PET and an as-grown $Sr_2IrO_4/Sr_3Al_2O_6$ layer on a $SrTiO_3$ substrate.] This material is available free of charge via the Internet at http://pubs.acs.org.

## AUTHOR INFORMATION

### Corresponding Author

**Ambrose Seo** – University of Kentucky, Lexington, Kentucky 40506, United States; https://orcid.org/0000-0002-7055-5314; Email: a.seo@uky.edu

### Notes

The authors declare no competing financial interest.

## ACKNOWLEDGMENTS

We acknowledge the support of a National Science Foundation grant DMR-1454200 for thin-film synthesis and characterization. M. Z. and J. H. acknowledge support by NSF, under Grant No. DMR-1847964. Electron microscopy was performed at the Center for Electron Microscopy and Analysis at The Ohio State University. This research used resources of the Advanced Photon Source, a U.S. Department of Energy (DOE) Office of Science User Facility operated for the DOE Office of Science by Argonne National Laboratory under Contract No. DE-AC02-06CH11357. A.S. acknowledges the support from the Alexander von Humboldt Foundation (Research Fellowship for Experienced Researchers) for Raman spectroscopy experiments.

Abstract Graphics

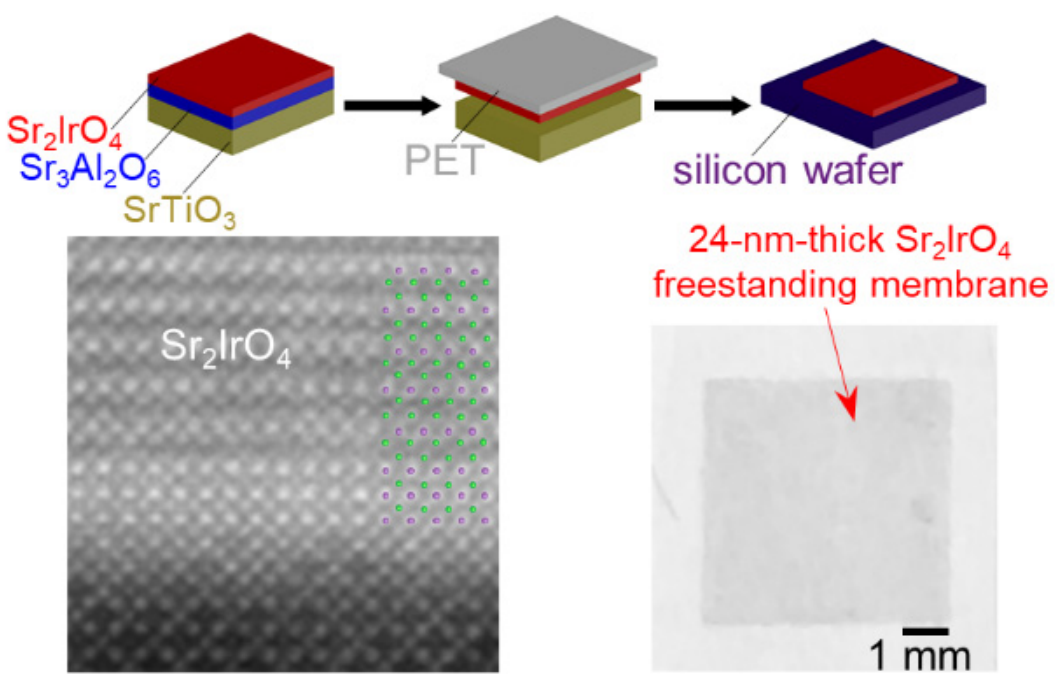